\documentclass[12pt,epsf]{article}
  \usepackage{epsfig}
  \newlength{\abstractwidth}
\newcommand{\onefigure}[2]{\begin{figure}[htb]
\begin{center}\leavevmode\epsfbox{#1.eps}\end{center}
\caption{#2\label{#1}}

 \end{figure}}

  \renewcommand{\thefootnote}{\fnsymbol{footnote}}
  \renewcommand{\thanks}[1]{\footnote{#1}} % Use this for footnotes
  \newcommand{\starttext}{
  \setcounter{footnote}{0}
  \renewcommand{\thefootnote}{\arabic{footnote}}}
  \renewcommand{\theequation}{\thesection.\arabic{equation}}
  \newcommand{\be}{\begin{equation}}
  \newcommand{\bea}{\begin{eqnarray}}
  \newcommand{\eea}{\end{eqnarray}}
  \newcommand{\beq}{\begin{equation}}
  \newcommand{\ee}{\end{equation}}
  \newcommand{\eeq}{\end{equation}}

  \def\ba{\begin{eqnarray}}
  \def\ea{\end{eqnarray}}

\def\14{{1\over4}}
  \def\12{{1 \over 2}}
  \def\eq{&=&}

  \def\h3{h^{3\over 2}}

  \def\tf{thermofield}
  \def\tfd{thermofield double}
  \def\lb{\label}

\begin{document}
  \renewcommand{\theequation}{\thesection.\arabic{equation}}
  \begin{titlepage}
  \bigskip

  \bigskip\bigskip\bigskip\bigskip

  \centerline{\Large \bf {1+1 Dimensional Compactifications of String Theory   }}

  \bigskip\bigskip
  \bigskip\bigskip
  %\centerline{\it }
  %\medskip
  %\centerline{} \centerline{} \centerline{}
  %\bigskip
   \begin{center}
  {\large Naureen Goheer }\\ Korea Institute for Advanced Study and Department of Mathematics and Applied Mathematics,
 University of Cape Town,\\7701 Rondebosch, Cape Town, South Africa\\
  \end{center}
   \begin{center}
  {\large Matthew Kleban }\\ Institute for Advanced Study, Princeton, NJ 08540, USA
  \end{center}
\begin{center}
  {\large Leonard Susskind}\\ Korea Institute for Advanced Study and Department of Physics,
  Stanford University\\ Stanford, CA 94305-4060, USA\\ \vspace{2cm}
  \end{center}

  \bigskip\bigskip
  \begin{abstract}

We argue that stable, maximally symmetric compactifications of string theory
to 1+1 dimensions are in conflict with holography.
In particular, the finite horizon entropies of the Rindler wedge in 1+1
dimensional Minkowski and anti de Sitter space, and of the de Sitter horizon in any dimension,
are inconsistent with the symmetries of these spaces.
The argument parallels one made recently by the same
authors, in which we
demonstrated the incompatibility  of the finiteness of the entropy
and the symmetries of de Sitter space in any dimension.  If the horizon entropy is either
infinite or zero the conflict is resolved.

  \medskip
  \noindent
  \end{abstract}

  \end{titlepage}
  \starttext \baselineskip=17.63pt \setcounter{footnote}{0}

%%%%%%%%%%%%%%%%

\setcounter{equation}{0}
\section{Introduction}

The holographic principle \cite{'thooft,world} has become one of
the most important ideas in quantum gravity.  It has led to many
major advances in the field in recent years. One may wonder
whether it constrains the possible backgrounds in string theory, or
any consistent theory that includes gravity. Indeed, in a recent
paper \cite{gks} we have argued that the symmetries of de Sitter
space are incompatible with the holographic requirement of finite
entropy of a causal patch. We further argued that the
incompatibility shows up in the very long time behavior of
inflating spaces, perhaps as an inevitable instability.

 In this note we will use the same argument
 to constrain the set of possible maximally symmetric compactifications of
string theory to two dimensions\footnote{We thank Juan Maldacena for suggesting the
application of the argument to AdS$_2$.}, at least in cases where they admit a geometric description. We will show that consistency with holography requires at least three space--time dimensions.

In \cite{gks}, the authors noted the fact that in all representations of the de Sitter group, the boost
generator (which generates time translations for the static patch) has a continuous spectrum.
However, the area of the de Sitter horizon is finite, which implies
a finite entropy for the quantum de Sitter Hamiltonian, and finite entropy is inconsistent
with a continuous spectrum.  Therefore, pure, eternal de Sitter space is inconsistent with holography.
The simplest explanation is that all de Sitter solutions in string theory are at best metastable,
with a finite lifetime followed by decay to a supersymmetric vacuum.

We can apply the same logic to 1+1 dimensional Lorentz and AdS invariant
compactifications of string theory.  The area
of the horizon of the Rindler wedge of 1+1 dimensional Minkowski space
is simply the volume of the compact internal space, and hence
is finite.  However, the 1+1 dimensional Poincar\'{e} algebra has the same commutation relation used in
\cite{gks} to prove continuity of the spectrum.  Therefore, by the same argument, the Rindler Hamiltonian
is continuous, which is inconsistent with the finite area of the horizon.  An identical argument can be made for 1+1 dimensional anti de Sitter space.  Our conclusion is
that maximally symmetric compactifications of string theory to 1+1 dimensions in which the horizon area is macroscopic are never stable.  One possible resolution is that the volume of the compact space is infinite; another is that it is very small and the entropy is zero.  In the latter case there is only order
one state in the Hilbert space, and the symmetry argument does not apply.

It is also the case that supersymmetric  compactifications to 1+1
dimensions will typically contain massless moduli, which in two dimensions
have an IR quantum instability. The expectation values of such
moduli will quickly spread to infinity ({\it e.g.} if an IR cutoff
is suddenly removed), which may account for the instability in the
size of the compact manifold.  Similar arguments were made in
\cite{bankssusskind}.

The plan of the paper is as follows:  in section 2, we review the formalism of thermofield
dynamics and its connection to the Rindler wedge of Minkowski space.  In section 3, we
prove (using the technique of \cite{gks}) that the Rindler Hamiltonian is continuous.  The same proof applies to the Hamiltonian of the ``Rindler" wedge of AdS.  In
section 4, we discuss the field theoretic instability, non-Lorentz invariant compactifications
to 1+1, and conclude.  A reader already familiar with the techniques of \cite{gks} may wish to skip
to section 4.

Note added:  after this paper was first posted on the arxiv, a preprint \cite{bdv} appeared which presented a supersymmetric compactification of heterotic string theory to 1+1 dimensions.  We believe that despite the lack of massless propogating degrees of freedom, due to IR divergences this model has a completely trivial S-matrix, and as such evades our argument in a way that will be mentioned briefly at the end of section 3.  This physics will be discussed further in \cite{mattsimeon}.

\section{Thermofield dynamics}

Thermofield theory was invented  \cite{tfd} in the context of many
body theory for the purpose of simplifying the calculation of real
time correlation functions in finite temperature systems. Its
connection with black holes was realized by Israel \cite{israel},
and elaborated in the holographic context by Maldacena
\cite{juan}.
In the thermofield formalism, one takes the tensor product of two
copies of the original field theory, labeled by $1,2$.  The two
copies are decoupled, and the total Hamiltonian is
\be
H_{tf} \equiv H \otimes I - I \otimes H, \ee where $H$ is the
Hamiltonian for the original theory and $I$ is the identity
operator. We then construct the entangled state \be |\psi \rangle=
{1 \over \sqrt{Z}} \sum_i e^{-\12\beta E_i}|E_i,E_i\rangle, \ee
where $|E_i,E_j\rangle=|E_i\rangle\otimes|E_j\rangle$, and
$|E_i\rangle$ are energy eigenstates. Strictly speaking the
correlation is not between identical states but between a state
and its CPT conjugate. The state $|\psi\rangle$ is a particular
eigenvector of $H_{tf}$ with eigenvalue zero. Furthermore
ambiguities in the construction due to degeneracies can be
resolved so that $|\psi \rangle$ is annihilated by the conserved
generators of symmetry transformations such as angular momentum or
electric charge.
Correlations between subsystems $1$ and $2$ are due to the
entanglement in $|\psi \rangle$.

Operators which belong to subsystem $1$ have the form $A \otimes
I$, and will be denoted
$A_1$. Operators associated with subsystem $2$ are defined in a
similar manner, except with an additional rule of hermitian
conjugation:
\be
A_2 \equiv I \otimes A^{\dag}. \ee Standard thermal correlation
functions may be written as  expectation values:
\be
\langle \psi | A_1(0) B_1(t) | \psi \rangle . \label{tfcora} \ee
As can be easily seen from the form of $ | \psi \rangle$,
\ref{tfcora} is simply the thermal expectation value of $A(0)
B(t)$, evaluated in a thermal density matrix at inverse
temperature $\beta$.  The state counting entropy observed in
subsystem $1$ is the entropy of entanglement of the state $|\psi
\rangle$.  In conventional applications, no physical significance
is usually attached to correlators involving both subsystems, but
we can certainly define them; for example

\be
\langle \psi| A_1(0) B_2(t) |\psi\rangle.
\label{tfcorb}
\ee
In the finite temperature AdS/CFT correspondence, \ref{tfcorb} has a
simple interpretation \cite{juan}: it corresponds to
a correlator  between operators on the two disconnected boundaries
of the spacetime.
It is not hard to see that one can compute \ref{tfcorb} by
analytically continuing \ref{tfcora}:
\be
\langle \psi| A_1(0) B_2(t) |\psi\rangle = \langle
\psi | A_1(0) B_1(-t-i\beta/2) | \psi \rangle .
\label{timecont}
\ee

\subsection{Rindler space}

We will now consider the relationship between thermofield dynamics
and quantum field theory in spaces with horizons. The simplest
example is Rindler space.
One plus one dimensional  Minkowski space
can be divided into four quadrants: I, II, III and IV (see Figure 1). Quadrants I and III
consist of points separated from the origin by a space--like
separation, while points in II and IV are displaced by timelike intervals.
Quadrant I is Rindler space, and can be described by
the metric
\be
ds^2=r^2 dt^2-dr^2,
\label{rind}
\ee
where $r$ is proper distance from the origin, and $t$ is
the dimensionless Rindler time. The Rindler quadrant may be described
by the Unruh thermal state with temperature
\be
T_{rind} ={1\over 2 \pi} ={1\over \beta_{rind}}
\label{trind}.
\ee

\onefigure{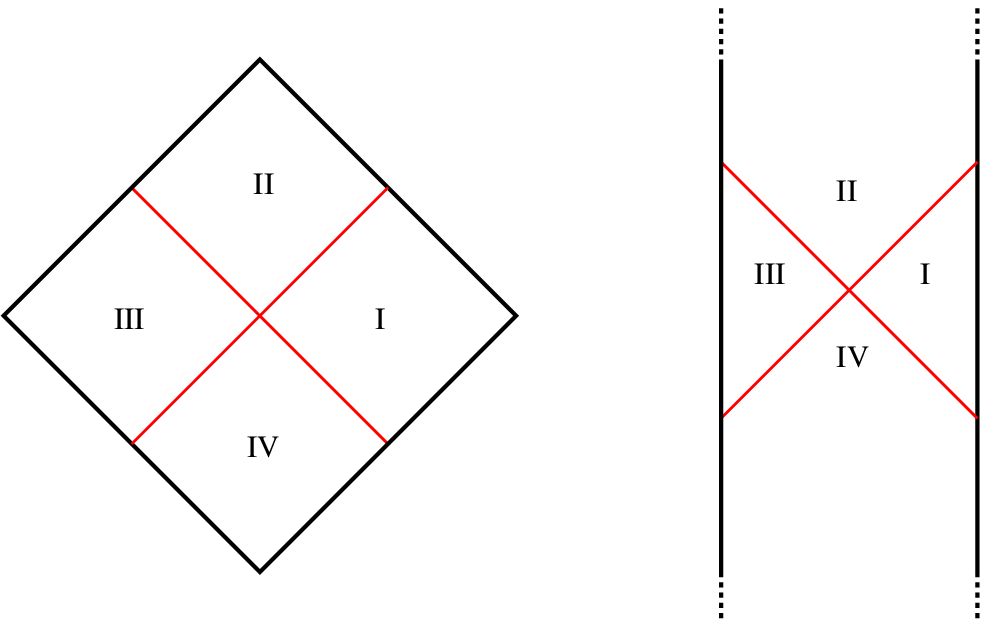}{On the left, the conformal diagram for Minkowski space,
showing the Rindler wedges I and III.  In Rindler coordinates, continuing
from region I to region III involves shifting time by $i \beta /2$,
in accord with \ref{timecont}. On the right, the ``Rindler" wedge of AdS$_2$.}

Quadrant III is a copy of quadrant I, and can be precisely
identified with the other half of the \tfd . To see this we first of all note that
the Rindler Hamiltonian is the boost generator. Since the
Minkowski vacuum is boost invariant, it is an eigenvector of the boost
generator with vanishing eigenvalue. Furthermore, the Minkowski vacuum
is an entangled state of the degrees of freedom in the two
quadrants I and III. Finally, it is well known that when the
density matrix for quadrant I is obtained by tracing over III
the result is a thermal state at the Rindler temperature.
It is easy to see that correlators between fields in
quadrants I and III are related by exactly the same analytic
continuations derived from \tf \ dynamics. To see this, recall that
the usual Minkowski variables $X^0,X^1$ are related to the Rindler
coordinates by
\bea
X^0 \eq r \sinh{ t} \cr
X^1 \eq r \cosh{t}.
\label{X}
\eea
Since the inverse  Rindler temperature \ref{trind} is $2\pi$, the
continuation in equation \ref{timecont} is
\be
t \to t-i\pi,
\label{cont}
\ee
or, from \ref{X}, $X^{\mu} \to -X^{\mu}$. Thus the \tf \
continuation takes quadrant I to quadrant III.

   In any dimension other than 1+1 the area of the horizon--and
   therefore the entropy of Rindler space--is infinite.
  However, if we
consider the space ${\cal R}_{1,1} \times {\cal M}_8$, where
${\cal M}_8$ is any eight-dimensional compact manifold, the area
of the Rindler horizon is nothing but ${\rm vol}({\cal M}_8)$, and
therefore  the entropy of the Rindler wedge must be finite.

A very similar construction produces the ``Rindler" wedge of anti de Sitter.  Start with the coordinates in which Euclidean AdS is a ball:
$ds^2 = d\rho^2 + sinh^2\rho d \Omega^2$.  Continuing the azimuthal angle of the sphere $\phi \rightarrow i t$
gives a metric covering one quarter of the AdS hyperboloid (see figure \ref{rindlerads}).  As in flat space, the horizon area is finite only for the case of AdS$_2$.

\section{Continuity of the spectrum}

It is a fact that any system with finite entropy must have a discrete spectrum;
actually, the statement is stronger: there must be a finite number of
states below any given energy $E$, so
the spectrum can not have discrete accumulation points.  Because this
is central to our analysis, we present a short proof below.

The entropy $S$ is defined by $S \equiv - {\rm Tr} \, \rho \ln \rho$, where
$\rho \equiv e^{- \beta H}/Z$ and $Z \equiv {\rm Tr} \, e^{-\beta H}$.  Here trace
means the sum over all the states in the spectrum of the Hamiltonian,
to be replaced by an integral with the proper measure
in the case that the
spectrum is continuous.  Expanding the relation above gives
\be
S = (1/Z) {\rm Tr}  \, \, e^{- \beta H} \left( \beta H + \ln Z \right)
= \ln Z + ( \beta/Z ) {\rm Tr} \, \,  H e^{- \beta H} ,
\label{entr}
\ee
which is nothing but the ordinary thermodynamic relation
$S = - \beta F + \beta E$.  As can be easily seen from \ref{entr}, $S$ is not affected
by the shift $H \rightarrow H - E_0$, for any constant $E_0$.  Therefore
we can take the ground state energy
to be zero (we will assume the energy is bounded from below).
Then the second term in \ref{entr} is manifestly positive, and
\be
S \geq \ln Z = \ln \sum  e^{- \beta H} \geq \ln \left( N e^{-\beta E_N} \right),
\ee
where $E_N$ is the $N$th energy level, and we have temporarily assumed that the
spectrum is countable.  Now the proof is clear:  if
there exists any energy $E_N$ below which there are an infinite number
of states, the entropy will diverge; and hence finite entropy implies
a discrete spectrum with no accumulation points.\footnote{Note that this does
{\em not} imply that the Hilbert space is finite.  In general, the
level spacing $\delta E \sim T e^{-S}$, at least for $T \gg \delta E$.}

This demonstrates that if the entropy of the Rindler wedge is
finite, then the Rindler Hamiltonian $H_R$ must have a discrete
spectrum.  The
implications for the thermofield Hamiltonian $H_{tf} $ are weaker.
The spectrum of this operator consists of the differences of
eigenvalues of $H_R$, namely $E_i-E_j$. If the spectrum of $H_R$ is
discrete, this set of numbers need not
be, but is certainly countable.  In the next section we will prove
that the spectrum of the full thermofield Hamiltonian $H_{tf}$
is not countable, and hence that the entropy of the thermal ensemble
defined using $H_R$ is infinite.

\subsection{Representations of the Poincar\'{e} algebra}

The Poincar\'{e} group in 1+1 dimensions has three generators,
which satisfy the following algebra: \bea &&[P_0,K] = iP_1 \cr
&&[P_1,K] = iP_0 \cr &&[P_0,P_1] = 0. \label{poingroup} \eea The
$P_i$ generate translations in space and time, and $K$ is the
boost generator. The choice of a Rindler wedge preserves only the
generator $K$ (this is obvious if one recalls that the Rindler
wedge is the region of space seen by a uniformly accelerating
observer).  The boost generator $K$ acts by generating
translations in Rindler time $t$; in other words, the Rindler
Hamiltonian is $ K$.  The generators $P_i$ do not act on the
Hilbert space of one wedge; rather, they act on the full
thermofield double space by mixing degrees of freedom  from the
two halves. In the case of a free field theory  this  means that
the modes of one Rindler wedge become a linear combination of
modes of both wedges under the action of the $P_i$; see the
appendix of \cite{gks} for more details.  The crucial point is
that the states of the thermofield double space, namely Minkowski
space, form a representation of the full group.

If we define $P_{\pm} = P_0 \pm P_1$, and identify the boost
$K$ with the thermofield Hamiltonian
$H$ the algebra becomes:\footnote{For simplicity of notation, 
for the rest of the section we will refer to the full
thermofield double Hamiltonian $H_{tf}$ simply as $H$.}

\bea &&[H, P_+] = -i P_+ \cr &&[H, P_-] = i P_- \cr
&&[P_+, P_-] = 0. \label{poingroup2} \eea

Using \ref{poingroup2}, we wish to prove  that the spectrum of
$H$ is continuous. Following \cite{gks}, consider the
operator $e^{i P_-}(t)$: \be e^{iP_- \, }(t) \equiv e^{iHt}
e^{i P_- \,} e^{-iH t} = e^{e^{iHt} \, i P_- \, \,
e^{-iH t}} =e^{iP_- \, \,  e^{-t}}. \label{Pt} \ee

We will now assume that the spectrum of $H$ is countable, and
use the assumption to derive a contradiction. We have \be \left|
\langle \alpha \,  | \, e^{i P_-  } \, | \,  \alpha \rangle
\right| = 1 - \delta. \label{lessthanone} \ee Here $| \alpha
\rangle$ is some state in the Hilbert space, and $\delta > 0$
because $e^{iP_-}$ is unitary and $P_-$ is non-zero. Define
\be
F(t) \equiv \langle \alpha \, | \, e^{iP_-(t)} \, | \,\alpha \rangle
= \langle \alpha \,|\,e^{iHt}e^{iP_- \,  }e^{-iHt}\,|\,\alpha \rangle
=\langle \alpha \,|\,e^{iP_- \,  e^{-t}}\,|\, \alpha \rangle.
\label{expec}
\ee
From this, $F(t) \rightarrow 1$ as
$t \rightarrow \infty$, and $F(0) = 1 - \delta < 1$.  We will now prove that
$F(t)$ is quasiperiodic (see e.g. the appendix of \cite{lisamatt}).

Any discrete sum of the form
\be
\sum_{n=1}^{\infty} f_n e^{i \omega_n t}
\ee
is quasiperiodic if
\be
\sum_{n=1}^{\infty} | f_n |^{\, 2} < \infty.
\label{criterion}
\ee
Therefore, it suffices to show that $F(t)$ can be written as a sum
of this form.
But, expanding the state $| \alpha \rangle$ in the energy 
basis:\footnote{The energies $\omega_i$ are the eigenvalues of the 
thermofield Hamiltonian $H = H_{tf}$, and hence are {\em differences}
of pairs of energies of the Rindler Hamiltonian $H_R$; but see the 
last paragraph before Section 3.1.}
\be
F(t) = \sum_{n,m} f^*_n f_m \langle n \, |\, e^{i P_-} \,|\, m \rangle e^{i (\omega_n - \omega_m)t}.
\ee
Consider the sum

\be
\sum_{m,n} f^*_n f_m f^*_m f_n  \langle n | e^{i P_-} | m \rangle
\langle m \,|\, e^{-i P_-} \,|\, n \rangle = \sum_n | f_n |^{\, 2}
\sum_m | f_m |^{\, 2} \langle n \,|\, e^{i P_-} \,|\, m \rangle
\langle m \,|\, e^{-i P_-} \,|\, n \rangle.
\ee
Considering the inner sum, we have (since $\sum_m  \langle n \,|\, e^{i P_-} \,|\, m \rangle
\langle m \,|\, e^{-i P_-} \,|\, n \rangle = 1$, and the terms are real and positive)
\be
\sum_m | f_m |^{\, 2} \langle n \,|\, e^{i P_-} \,|\, m \rangle
\langle m \,|\, e^{-i P_-} \,|\, n \rangle \leq 1,
\ee
and therefore
\be
\sum_{m,n} f^*_n f_m f^*_m f_n  \langle n \,|\, e^{i P_-} \,|\, m \rangle
\langle m \,|\, e^{-i P_-} \,|\, n \rangle \leq 1.
\ee
This shows that $F(t)$ satisfies the criterion \ref{criterion}, and hence
$F(t)$ is quasiperiodic.  Therefore, since $F(0) < 1$, $F(t)$ can not tend to $1$ as
$t \rightarrow \infty$, and we have a contradiction.

We can make almost exactly the same argument using the ``Rindler" wedge of AdS$_2$.  The algebra
is $SO(2,1)$ (which is identical to the dS$_2$ algebra) and in fact the analogy to
the argument of \cite{gks} is exact.  The generator of ``Rindler" time translations is
one of the boost generators of the AdS hyperboloid.  Again, the argument shows
the the spectrum must be continuous.

This proves that $H_{tf}$ can not have a countable spectrum, and
therefore that the $E_i$ cannot be discrete and the entropy cannot
be finite.  However, the area of the Rindler horizon in 1+1
dimensions is finite, and so we see there is a fundamental
conflict between the holographic principle and the existence of a
stable compactification of string theory to 1+1 dimensional
Minkowski, AdS, or dS space.

We note one possible loophole--if the representation is trivial, so that there are only vacua in the spectrum, our argument fails.  In that case the generators are
zero, and there is no conflict with the algebra.  We believe the examples presented in \cite{bdv} fall into this category, as any state with non-zero energy will cause an infinite back-reaction on the geometry.

\section{String theory arguments}

Let us begin with supersymmetric compactifications of string
theory. In general there will always be massless moduli such as
the overall size of the compact space. These massless degrees of
freedom can be thought of as $1+1$ dimensional scalar fields. But
it is well known that in $1+1$ dimensions there are logarithmic
infrared divergences. For example the two point function has a
divergence
 of the form $\int dk/k$.  While the UV
divergence can of course be regulated, the IR divergence
represents a true physical effect; namely that the fields
fluctuate more and more at longer and longer wavelengths. In
particular, if the Hamiltonian includes a mass term $m^2
 \, \theta(-t)$, where $\theta$ is the unit step function, the
expectation value $\langle \phi^2(t) \rangle$ will tend to
infinity like $t$, for $t>0$.  This means that the field, once released from
its confining potential, fluctuates more and more in field space.
This effect indicates a decompactification of the manifold. In
other words the compactification is unstable.

This argument (which was discussed previously in \cite{bankssusskind})
is in some ways more generic than the argument presented above, because
it does not rely on maximal symmetry.  On the other hand, it does require
that at least some of the compactification moduli are massless.

There are, of course, $1+1$ dimensional compactifications of
string theory such as the linear dilaton vacua in which the
dilaton varies linearly, either with respect to time or space.
However, these obviously violate Lorentz invariance. What is more,
at the weak coupling end the volume of the compact space diverges
if expressed in Planck units. Other possible examples, such as pp
wave in anti de Sitter space, can be thought of as Lorentz
non--invariant $1+1$ dimensional theories and do not contradict
our conclusion.

\section{Conclusion}

In both \cite{gks} and the present paper we have found that the
delicate symmetries that ensure the equivalence of different
observers (observer compementarity ) can not be implemented for
systems of finite entropy. It seems that for these symmetries to
be exact, an infinite horizon area must be available for
information to spread out in. We think that this is a general
rule: Exact observer complementarity is only possible if the
horizon is infinite in extent.

This raises the question of finite mass black hole horizons. In
this case there is no exact symmetry between observers outside the
black hole and those which fall through the horizon. Indeed
sufficiently careful measurements of tidal forces should be able
to tell a freely falling observer exactly when she crosses the
horizon. Only in the limit of infinite mass does the horizon
become precisely undetectable.

\section{Acknowledgements}
The authors would like to thank C. W. Kim and the faculty and staff of KIAS for their hospitality, where this work
was initiated. We thank Jacques Distler, Simeon Hellerman, and Juan Maldacena for discussions.  N.G. would like to thank UCT for supporting her
studies there.

The work of M.K. is supported by NSF grant PHY-0070928, and that
of L.S. by NSF grant PHY-0097915.

\end{document}